\documentclass[singlespacing]{elsart}

\usepackage{graphicx,natbib,amssymb,color,lineno}

\begin{document}

\begin{frontmatter}

\title{Comparison between Eulerian diagnostics and finite-size Lyapunov
exponents computed from altimetry in the Algerian basin}

\author[lmd]{F. d'Ovidio\corauthref{cor}}

\corauth[cor]{Corresponding author.}
\ead{dovidio@lmd.ens.fr}

\author[brest1,brest2]{Jordi Isern-Fontanet}
\ead{jisern@ifremer.fr}

\author[ifisc]{Crist\'obal L\'opez}
\ead{clopez@ifisc.uib.es}

\author[ifisc]{Emilio Hern\'andez-Garc\'\i a}
\ead{emilio@ifisc.uib.es}

\author[icm]{Emilio Garc\'{\i}a-Ladona}
\ead{emilio@cmima.csic.es}

\address[lmd]{Laboratoire d'Oc\'eanographie et du Climat: Exp\'erimentation et Approches
Num\'eriques, IPSL, Paris, France and Institute of Complex Systems, Paris
Ile-de-France (ISC-PIF)}

\address[brest1]{Laboratoire de Physique des Oc\'eans, Ifremer, Plouzan\'e, France.}

\address[brest2]{Laboratoire d'Oc\'eanographie Spatiale, Ifremer, Plouzan\'e,
France.}

\address[ifisc]{Instituto de F{\'\i}sica Interdisciplinar y Sistemas Complejos IFISC 
(CSIC-UIB),
  Campus Universitat de les Illes Balears, E-07122 Palma de Mallorca, Spain.}

\address[icm]{Institut de Ci\`encies del Mar,
Passeig Mar\'{\i}tim de la Barceloneta 37-43, E-08003 Barcelona, Spain.}


\begin{abstract} Transport and mixing properties of surface currents can be
detected from altimetric data by both Eulerian and Lagrangian diagnostics. In
contrast with Eulerian diagnostics, Lagrangian tools like the local Lyapunov
exponents have the advantage of exploiting both spatial and temporal variability
of the velocity field and are in principle able to unveil subgrid filaments
generated by chaotic stirring. However, one may wonder whether this theoretical
advantage is of practical interest in real-data, mesoscale and submesoscale
analysis, because of the uncertainties and resolution of altimetric products,
and the non-passive nature of biogeochemical tracers. Here we compare the
ability of standard Eulerian diagnostics and the finite-size Lyapunov exponent
in detecting instantaneaous and climatological transport and mixing properties.
By comparing with sea-surface temperature patterns, we find that the two
diagnostics provide similar results for slowly evolving eddies like the first
Alboran gyre. However, the Lyapunov exponent is also able to predict the
(sub-)mesoscale filamentary process occuring along the Algerian current and
above the Balearic Abyssal Plain. Such filaments are also observed, with some
mismatch, in sea-surface temperature patterns. Climatologies of  Lyapunov
exponents do not show any compact relation with other Eulerian diagnostics,
unveiling a different structure even at the basin scale. We conclude that
filamentation dynamics can be detected by reprocessing available altimetric data
with Lagrangian tools, giving insight into (sub-)mesoscale stirring processes
relevant to tracer observations and complementing traditional Eulerian
diagnostics.

 \end{abstract}


\begin{keyword}
Submesoscale \sep filaments \sep altimetry \sep Mediterranean circulation \sep
Lagrangian dynamics
\end{keyword}
\end{frontmatter}
\linenumbers

\fontsize{10}{12}
\selectfont

\section{Introduction}

Satellite high-resolution daily images of tracers like sea-surface temperature
and chlorophyll show a large mesoscale and sub-mesoscale heterogeneity and
patchiness, typically in the form of filaments. The full range of tracer
variability observed in a climatological mean at the basin scale, can occur in
daily snapshots over spatial scales of just 10-100 km (see for instance
\citet{Lehahn.2007} for the case of chlorophyll in the NE Atlantic). This large
hetereogenity occurring on relatively short distances is able to induce very
strong tracer gradients and to impact on important aspects of the ocean dynamics
like lateral transport, upwelling/downwelling, and mixing. Mesoscale and
sub-mesoscale variability is an important component of plankton dynamics
\citep{Abraham.1998,PCE.2000, abraham.2000, Boyd.2000, Toner.2003, Martin.2003},
larval transport \citep{Bradbury.2001}, as well as the dispersion of
contaminants and oil spills. In polar regions, (sub-)mesoscale tracer
variability has been recently recognised to impact also on the thermohaline
properties of the mixed layer, affecting the low frequency ocean circulation
through the formation of waters below the mixed layer \citep{Sallee.2006a, Sallee.2007b}.
Mesoscale and sub-mesoscale gradients of heat and salinity can induce cells of
very strong vertical velocities (several tens of meters per day) from subsurface
down to and below the mixed layer depth. These vertical velocities may fertilise
the photic layer, creating local plankton blooms and affecting the sequestration
of organic matter from the surface \citep{Levy.2001, Levy.2005a}. This mechanism
can affect the properties of deep waters when they are formed \citep{Paci.2005}.
Tracer gradient intensification is also a precondition to mixing. 
Horizontally, the elongation of water masses in thin structures intensifies
local gradients, greatly enhancing background diffusion \citep{Lapeyre.2006}.
Vertically, the trapping of inertial waves inside filaments creates mixing
hotspots that extends to the bottom and below the mixed layer depth
\citep{Young.1982}.

Tracer mesoscale and sub-mesoscale patches are often found associated to
surface mesoscale eddies (e.g. \citet{Robinson.1993, McGillicuddy.1998,
Martin.2002, Abraham.2002, Morrow.2004, Legal.2006}). Tracers are stretched into filaments by the
shear-dominated regions in between mesoscale eddies while the
recirculating regions inside eddies' cores can trap and transport tracer
anomalies for timescales comparable to the eddy lifetime. As a well-known tool
to reveal geostrophic velocities associated to eddies,
altimetry data when opportunly analyzed, provide an important source of
information for the mesoscale and sub-mesoscale structuration of tracer
distributions.

Traditionally, transport information has been inferred from altimetry with
Eulerian diagnostics like the eddy kinetic energy, the Okubo-Weiss criterion
that separates the eddy's core from the periphery and others, that will be
discussed in detail shortly. These diagnostics are based on the analysis of
instantaneous snapshots of tracer and velocity fields. For instance, the
Okubo-Weiss (OW) parameter has been applied to drifters in order to characterize
eddy properties \citep{Stocker.2003,Testor.2005}, to in situ density profiles to
identify regions with different mixing properties \citep{Isern.2004}, and to
altimetric maps with similar objectives \citep{Isern.2004,Waugh.2006}. The OW
parameter has also widely been used for the identification and characterization
of ocean eddies from altimetry
\citep{Isern.2003,Morrow.2004,Chaigneau.2005,Isern.2006}. More recently, it has
been suggested that Lagrangian diagnostics are more appropriate to link
turbulence properties to tracer dynamics. The most used Lagrangian diagnostics
have been the calculation of local Lyapunov exponents, that measure the relative
dispersion of advected particles.  In contrast to Eulerian tools, Lagrangian
diagnostics do not analyze instantaneous snapshots of the velocity field, but
measure transport properties along particle trajectories, therefore
reconstructing the fine structure of transport dynamics that a fluid parcel has
experienced, like subgrid filament formation. These patterns depend on the
advection history along a trajectory, that may span a large spatiotemporal domain
of the velocity field, and therefore cannot be captured by Eulerian diagnostics,
that instead measure local, properties. Regarding Lagrangian tools
like the finite-size or finite-time Lyapunov exponents (resp. FSLEs and FTLEs)
\citet{Abraham.2002} computed FTLEs from time-dependent velocity fields and
compared to istantaneous SST patterns, \citet{LaCasce.2003} used FSLEs to study
the statistics of a large data set of drifting buoys; \citet{Waugh.2006} applied
FTLEs to altimetry and compared the results with OW and kinetic energy
estimations; \citet{Lehahn.2007} computed FSLEs from the geostrophic field in
order to extract transport barriers and compared with satellite derived plankton
and SST patches; \citet{Rossi.2008} compared mixing and productivity by using the
FSLE. Besides these studies, the FSLE has been also applied to numerical
simulations of the Mediterranean sea with different objectives: the
predictability of Lagrangian trajectories \citep{Iudicone.2002} and the
characterization of flow patterns \citep{dOvidio.2004} and dispersion
\citep{Garcia-Olivares.2005}.

However, for the case of real velocity data and given the chaotic nature of
mesoscale advection (in the sense of positive Lyapunov exponents) one may expect
Lagrangian diagnostics to be more affected than Eulerian tools by any space and
time indetermination of the velocity field. When dealing with real altimetric
data, therefore one may ask whether the theoretical advantage of Lagrangian
techniques of detecting small scale transport structure is not countered in
practice by the intrinsic error and cutoff of altimetric data, so that this
information is either unreliable or already contained in Eulerian quantities.
This question is particularly relevant given the larger computational costs (a
few orders of magnitude) of Lagrangian tools in respect to Eulerian ones. In
fact, for the case of the Tasman sea \citet{Waugh.2006} found a striking
resemblance between maps of Lyapunov exponents and the eddy kinetic energy, as
well as a compact relationship between the two quantities. In their concluding
remarks, \citet{Waugh.2006} argued that this observation, if valid in other
oceanic basins as well, would rise the possibility of using the EKE for
estimating stirring rates, without the need of explicit, longer Lagrangian
calculations. However, one of the main advantages of the Lagrangian techniques
is the possibility of reconstructing tracer patterns that are below the
resolution of the velocity field and that arise due to several iterations of
stretching and folding during the tracer (chaotic) advection. Indeed, in
\citet{Waugh.2006} the correlation between the Lyapunov exponents have been
found on a grid at relatively low resolution (0.5 deg.) and for a advection
times of 14 days (two altimetric images), i.e. without the spatial and temporal
information needed for the developement of filaments. Does this result hold also
for filament-resolving Lagrangian calculations?

In this paper we address this question, exploring the possibility of obtaining
reliable information with Lyapunov exponents that is not already contained in
Eulerian diagnostics, focusing on filament dynamics.

This will be done both on the analysis of individual patterns (validated by high
resolution SST images) and on a climatological basis. We find that at
(sub-)mesoscale resolution the information provided by Eulerian diagnostics and
Lyapunov exponents coincindes only for very stationary eddies, while providing
two distinct and complementary pictures of the circulation in all the other
cases: the Eulerian analysis provides the eddies that populate the mesoscale,
while the Lagrangian analysis yields the tracer filaments generated by the
spatiotemporal variability of these eddies. Our results also show a surprising
reliability of altimetric data at the scale of their nominal resolution, when
reprocessed with Lagrangian tools.

We will compare Eulerian diagnostics and Lyapunov exponents in the Algerian
basin because of its rich and variable mesoscale activity, that contains
jet-dominated regions (the path of the Algerian current) as well as eddies with
very different characteristics (from the quasi-stationary Alboran gyres, to the
slowly propagating eddies of the Balearic Abyssal plain, and the fast-evolving
Algerian eddies). A main characteristic of the surface circulation in the
Mediterranean sea is the propagation of fresh waters incoming from the Atlantic
ocean. At the entrance of the Mediterranean, these waters flow from west to east
and form patterns such as the Alboran eddies east of the Gibraltar strait or the
Algerian current along the Algerian coast. The instabilities of this current
generates, a few times per year, coastal eddies that propagate downstream,
usually until the entrance of the Sardinia channel. There, they can detach from
the coast and propagate as open sea eddies \citep{Millot.1999} following
relatively well defined paths \citep{Isern.2006}. These eddies, called Algerian
eddies, have variable diameters of about $50-200\;km$, vertical extents from
hundreds to thousands of meters, and lifetimes of several months, up to nearly 3
$years$ \citep{Millot.1997,Puillat.2002}. Their presence has a large impact on
the redistribution of tracers in the Algerian basin, which is characterized by
the northward spreading of tracers that are initially transported eastwards by
the  Algerian current \citep{Ovchinikov.1966,Brasseur.1996}. As we shall see,
our analysis of stirring will unveil the crucial role on tracer patterns of the
time variability of the mesoscale activity in the Algerian basin and the
relevance of topography in constraining the dynamics of coherent structures.

The paper is organized as follows. After describing the data sets and the
techniques, we develop an Eulerian and Lagrangian analysis for specific days. We
focus on the representation of eddies and the detection of transport barriers
and we get two complementary pictures: a regular and smooth OW-based description
and a more complex, lobular representation from the FSLE. We then compare such
different structures to tracer distributions. In order to filter out the active
dynamics of a real tracer as well as the unresolved components and
indeterminacies of altimetry data, we first consider the filaments of a
synthetic tracer, that we advect numerically with the altimetric data. As a
second step, we take sea-surface temperature (SST) satellite images. The
indications that are found on the individual days are then generalized in a
climatological comparison, where we compare the spatial variability of FSLE, OW
parameter, strain rate, and eddy kinetic energy temporally averaged over the
period 1994-2004. For both the Eulerian and the Lagrangian
analysis, we also propose to describe the consequences of filament formation
with a climatology giving the spatial density of transport barriers.

\section{Methods}\label{sec:method}

A traditional approach to the characterization of the stirring and mixing in the
presence of mesoscale eddies consists in separating the stagnation region at
the eddy's core from the eddy's periphery where tracer filamentation occurrs.
This is done by measuring the relative dominance of vorticity and deformation.
One of the most used parameters for measuring this relative dominance is the
Okubo-Weiss parameter \citep{Okubo.1970,Weiss.1991} which has been already used
by some of us to study  properties of Algerian eddies \citep{Isern.2004}. The OW
parameter is a particular case of the more general vortex-identification
criterion proposed by \citet{Jeong.1995}.

The Okubo-Weiss parameter $W$ is defined as:

\begin{equation} W=s_n^2+s_s^2-\omega^2 \label{OWp} \end{equation}

where $s_n$, $s_s$ and $\omega$ are the normal and the shear components of
strain, and the relative vorticity of the flow defined, respectively, by \[
s_n=\frac{\partial u}{\partial x}-\frac{\partial v}{\partial y}, \quad
s_s=\frac{\partial v}{\partial x}+\frac{\partial u}{\partial y}, \quad
\omega=\frac{\partial v}{\partial x}-\frac{\partial u}{\partial y}. \]

In the formula above, $x$ and $y$ are orthogonal spatial coordinates and $u$ and
$v$ are the component of the velocity respectively for the $x$ and $y$
directions. 

The parameter $W$, allows to separate a two-dimensional flow into different
regions: a vorticity-dominated region ($W < -W_0$), a strain-dominated region
($W > W_0$) and a background field with small positive and negative values of
$W$ ($|W| \leq W_0$). Here $W_0=0.2\sigma_W$, $\sigma_W$ being the standard
deviation of the $W$ values in the whole domain, in our case the Mediterranean
sea \citep{Bracco.2000b,Pasquero.2001,Isern.2006, Elhmaidi.1993}. The core edge
can then be identified as the closed lines with $W=0$. This separation of the
field in terms of the sign of $W$ has been proved to be a robust criterion for
extracting eddy cores from complex fluid flows \citep{Jeong.1995,Pasquero.2001}.
In steady flows, the boundary of the core constitutes a barrier to the exchange
of particles with the surrounding cell, so that particles trapped inside the
eddy core remain there. Under non-steady flows, particles can leak out of the
vortices, be ejected through filamentation processes, or even the eddy can be
destroyed \citep{Basdevant.1994,Hua.1998}.

 The OW parameter has some well-known limitations. On one side, it assumes that
velocity gradients are slowly evolving in time, which is only valid inside
relatively coherent regions. On the other side, this parameter
fails to properly identify regions with different mixing properties when eddies
are stationary and have axial symmetry \citep{Lapeyre.1999}. There have been
some attempts to solve some of these limitations by extending it to consider the
time evolution of velocity gradients \citep{Hua.1998,Hua.1998b, Lapeyre.1999}.

Another approach to the characterization of flow structures is to assume a
Lagrangian viewpoint, that is, to look explicitly at transport properties from
the analysis of particle trajectories \citep{Pierrehumbert.1991a,
Ridderinkhof.1992,Miller.1997,Haller.2000, Coulliette.2001,
Koh.2002,dOvidio.2004, Shadden.2005, Mancho.2006}. In contrast to the OW method
and in general to Eulerian diagnostics, this approach requires knowledge of the
time variability of the velocity field, as well as the use of an integrator for
generating the trajectories. Several methods exist, one of the simplest being
the Finite-Size Lyapunov Exponent (FSLE). The FSLE is a generalization of the
concept of Lyapunov exponent to finite separations. The standard definition of 
Lyapunov exponent refers to  the exponential rate of divergence, averaged over
infinite time, of infinitesimally closed initial points. The FSLE technique
keeps the original idea of capturing the rate of divergence between
trajectories, but overcomes the limit operations. Thus, it is (and has been
shown to be) rather appropriate to manage real data. The FSLEs were introduced
for turbulent flows \citep{Aurell.1997,Artale.1997} aiming at studying
non-asymptotic dispersion processes. Since then, they have been used for two
complementary goals: for characterizing dispersion processes
\citep{Lacorata.2001}, and for detecting and visualizing Lagrangian structures
(e.g.  transport barriers and fronts) \citep{Boffetta.2001,
Koh.2002,Joseph.2002,dOvidio.2004}. In the framework of this paper, in order to
compare with the OW parameter we will focus mainly on the second use.

Several methods allow to calculate the FSLEs. In the simplest
scheme, for each instant $t$ and each point $\bf x$, one follows
in time the evolution of a tracer started in $\bf x$ and of
another probing tracer located at a distance $\delta_0$ from it.
The integration is stopped when the two tracers have reached a
final separation $\delta_f>\delta_0$. From the time interval,
$\tau$, to reach the final separation, the FSLE is defined in the
following way:

\begin{equation}
\lambda ({\bf x},t, \delta_0,\delta_f) =
\frac{1}{\tau}\log\frac{\delta_f}{\delta_0}.
\label{eq:fsle}
\end{equation}

\noindent

In order to reduce the dependence on the direction of the probing tracer, the
algorithm is run choosing three points forming an equilateral triangle around
$\bf x$. We stopped the integration when any of these three points reaches a
separation $\delta_f$ from the trajectory started in $\bf x$.

\begin{figure}
\begin{center}
\includegraphics[width=19pc]{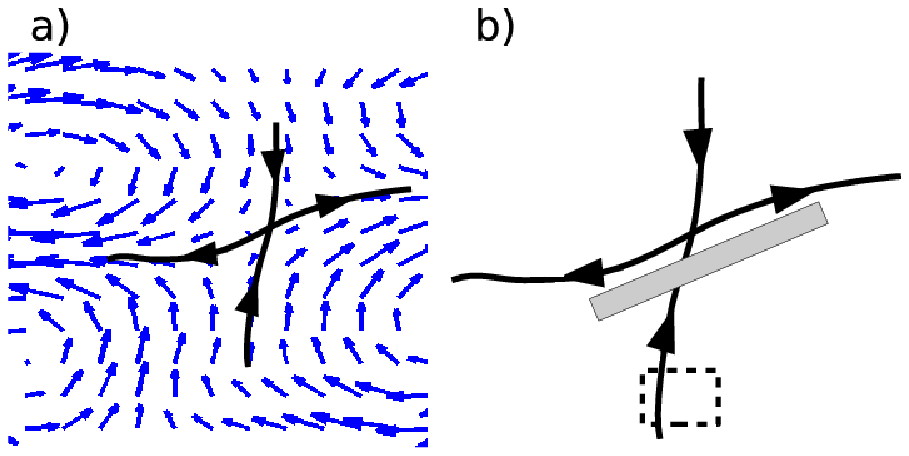}
\end{center}
\caption{A sketch of a hyperbolic structure (a) and its effect on an tracer
advected by the velocity field and initialised in the dashed region (b).
The unstable manifold acts as a transport barrier and generates a
front, while the stable manifold cannot be directly compared with
the tracer distribution. A more quantitative picture is presented
in Fig.~\ref{fig:tracer}b. \label{fig:saddle}}
\end{figure}

Maxima (ridges) of Lyapunov values are typically organized in convoluted, lobular
lines. For the case of the backward calculations, these lines can be interpreted
as the fronts of passively advected tracers. This interpretation can be
understood in a very qualitative but effective way, considering that fronts
typically separate fluid patches of different origins. Therefore, the separation
in the past of a couple of points is largest when the couple is initialised exactly
over the front. This argument can be rephrased in a slightly more rigorous way in
the context of dynamical systems, interpreting the line-shaped regions of fastest
separation -either backward or forward in time- respectively as the unstable and
stable manifolds of the hyperbolic points in the flow
\citep{Haller.2000,Boffetta.2001,Haller.2001a, Koh.2002,Joseph.2002,dOvidio.2004,
Mancho.2006}. The effect on advection of hyperbolic structures is sketched in
Fig.~\ref{fig:saddle}. Due to the convergent field along the stable manifold and
the divergent field along the unstable manifold, a passively advected tracer is
deformed as in Fig.~\ref{fig:saddle}, developing a front along the unstable
manifold and a gradient orthogonally to it. Due to the hyperbolic structure of
which the manifold is a part, the tracer front approaches the manifold
exponentially fast. The distance $\delta_r$ between the tracer front and the
manifold depends on the initial front-to-manifold distance $\delta_i$, the
exponent $\lambda$ of the manifold, and the time of integration $t$:

\begin{equation}\label{eq:relax}
\delta_r\approx \delta_i \exp(-\lambda t).
\end{equation}

Manifolds characterized by higher exponents have therefore a stronger effect on
tracers, shaping a front in shorter times and being more visible in
tracer distributions.

Note that for a time-dependent velocity field, the sketch of
Fig.~\ref{fig:saddle} holds only if the hyperbolic structures evolve in time on
a time scale slower than the tracer advection, so that the tracer can actually
relax over the manifold. See \citet{Lehahn.2007} for more details on the FSLE
computation from altimetric data.

Mixing properties can also be diagnosed by Lyapunov exponent calculations,
either considering the exponential separation in the future as a measure of
tracer dispersion, or by combining forward and backward information
\citep{dOvidio.2004}. Since in this work we aim at a direct comparison with
advected tracers (SST), we will focus on the backward calculation and compare
the location of the manifolds detected by the FSLEs with tracer fronts.

As it is clear from Eq.(\ref{eq:fsle}), the FSLEs depend critically on the
choice of two length scales: the initial separation $\delta_0$ and the final one
$\delta_f$. \citet{dOvidio.2004} argued that $\delta_0$ has to be close to the
intergrid spacing $\Delta x$ among the points $\bf x$ on which the FSLEs will be
computed. In fact, a $\delta_0$ larger than intergrid spacing would allow to
sample manifolds of strong Lyapunov exponents in more than one grid points,
while $\delta_0$ smaller than the intergrid spacing would not allow to follow a
manifold on the sampling grid as a continuous line. Following \citet{dOvidio.2004,Lehahn.2007}, in this work
we have set $\delta_0=0.01^o$ (approx. 1 km) in order to match the resolution of
SST images. We have set  $\delta_f=1^o$  i.e., separations of about $110$ Km,
that is the order of magnitude of the eddies' radii detected by altimetry.
Values of $\delta_f$ smaller or larger up to 50\% do not change significatively
the calculation. The time of integration for finite-size Lyapunov exponents
varies from one point of another, being small for strong values and vice versa.
Inverting Eq.\ref{eq:fsle},

\begin{equation}
\tau =
\frac{1}{\lambda}\log\frac{\delta_f}{\delta_0}.
\label{fsletau}
\end{equation}

We found typical values of finite-size Lyapunov exponents in the range $0.1-0.2$
days$^{-1}$, corresponding to integration times of resp. 46 and 23 days.

\section{Data}

\subsection{Sea-Surface Height}

In the Mediterranean sea, despite the weak signal intensity and
the coarse space and time resolution of the altimetric tracks,
several studies have shown the reliability of the altimetric data
to analyze its dynamics, particularly in the Algerian basin
\citep{Vignudelli.1997, Bouzinac.1998,Larnicol.2002,Font.2004}. In
this study we have  used Delayed Time Maps of Absolute Dynamic
Heights (DT-MADT) produced by {\it Collecte Localisation
Satellites  (CLS)} in Toulouse (France) specifically for the
Mediterranean sea, which combine the signals of {\em ERS-ENVISAT}
and {\em TOPEX/Poseidon-JASON} altimeters.

Altimetric data are processed including usual corrections
(sea-state bias, tides,  inverse barometer, etc.) and improved
orbits. From several corrected sea-surface height files, a
conventional repeat-track analysis  is performed to extract the
Sea Level Anomaly (SLA) relative to a mean profile:  data are
re-sampled along the mean profile using cubic splines and
differences relative to the mean profile are calculated.  SLA
along-track data are then filtered and subsampled. The filters
used  are a non linear median over 3 points (roughly 21 km)
followed by a low pass along  track linear Lanczos filter (with a
cut-off wavelength of 42 km). SLA data are then  subsampled every
other point \citep{AVISO.2006}. Finally, SLA maps are built using
an improved space/time objective analysis method, which takes into
account long wavelength  errors, on a regular grid
\citep{LeTraon.1998b} of $(1/8)^o\times(1/8)^o$ every week. Then,
Sea-Surface Heights (SSH) are finally obtained by adding to the
SLA a Mean Dynamic Topography \citep{Rio.2007}.

For each data set geostrophic velocities are estimated as usual: \begin{equation}
u=-\frac{g}{fR_T}\frac{\partial h_{ssh}}{\partial \phi},\;
v=\frac{g}{fR_T\cos\phi}\frac{\partial h_{ssh}}{\partial \lambda}, \label{GEOeq}
\end{equation} where $h_{ssh}$ is the SSH, g is gravity, $f$ the Coriolis
parameter, $R_T$ the Earth radius, $\phi$ the latitude and $\lambda$ the
longitude. The data analyzed spans from January 1, 1994 to December 31, 2004.
From this data set we study representative days and we also construct
climatologies. Finally, a Runge-Kutta integrator of fourth order and a time step
of 6 hours has been used to obtain backward and forward trajectories in the
velocity field for the calculation of FSLEs and for the advection of a synthetic
tracer. The geostrophic velocity field has been resampled in space and time with
a multilinear interpolator.

\subsection{Sea-surface temperature}

We used sea-surface temperature (SST) data from the AVHRR sensors on board NOAA
satellites, downloaded from the HRPT station at the Institut de Ci\`encies del
Mar (CSIC) in Barcelona. For the single day analysis we have chosen images for
which there were good quality sea-surface temperature images presenting a large
variety of eddy structures and different intensities of the Algerian current:
July 9, 2003; April 7, 2004; June 30, 2004. SST images have a resolution of 1.1
km at the nadir.

\section{Results}

\subsection{Eddy representation}

\begin{figure}
\begin{center}
\includegraphics[width=39pc]{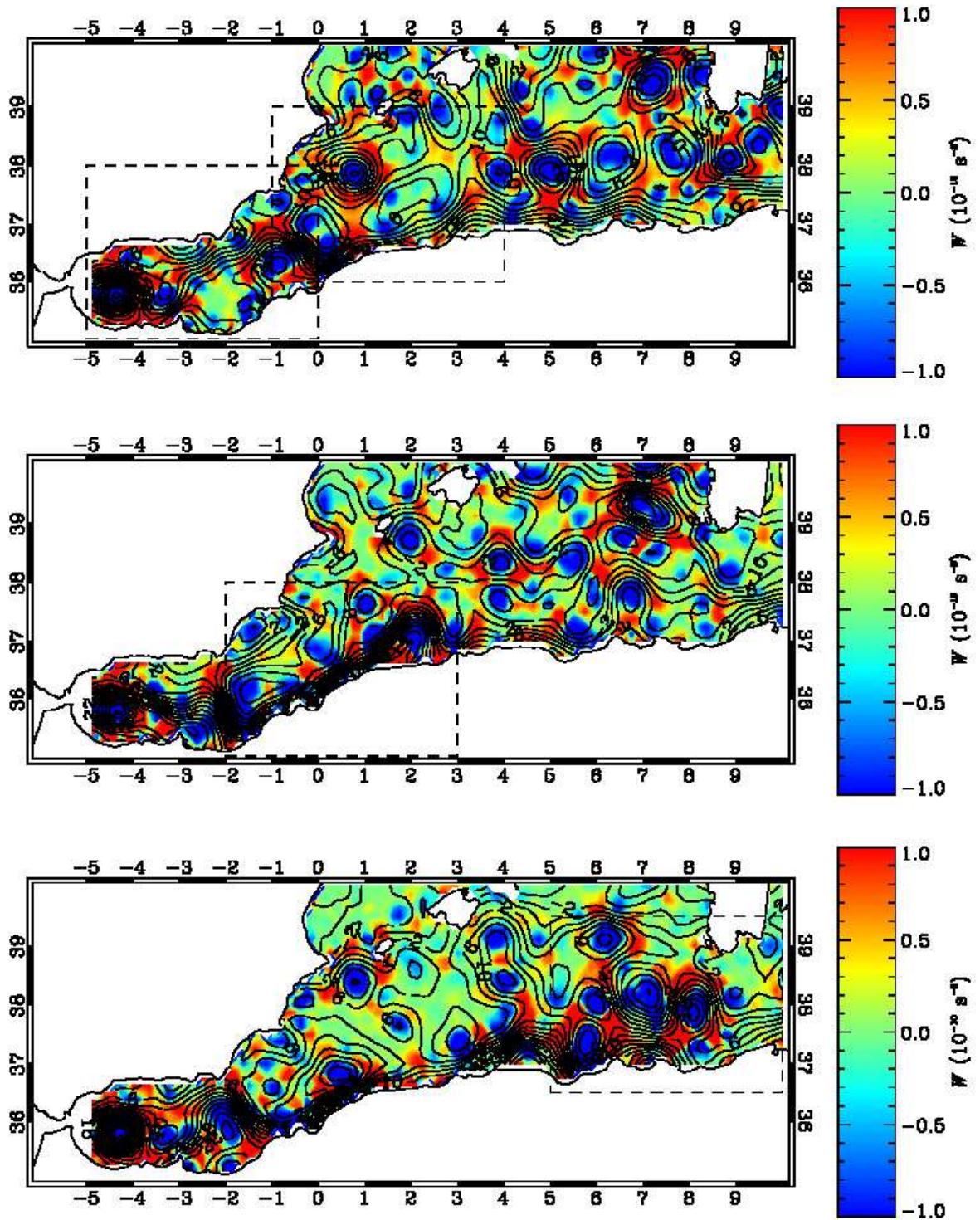} \\
\end{center}
\caption{Okubo-Weiss parameter ($W$, coded in colors) computed from sea-surface
height
(black line) fields
corresponding (from top to bottom) to July 9, 2003; April 7, 2004;
June 30, 2004. The dashed boxes represent the regions over which we perform a
comparison with SST (see Figs.~\ref{fig:sst} and ~\ref{fig:tracers}).
\label{fig:okw}}
\end{figure}

\begin{figure}
\begin{center}
\includegraphics[width=39pc]{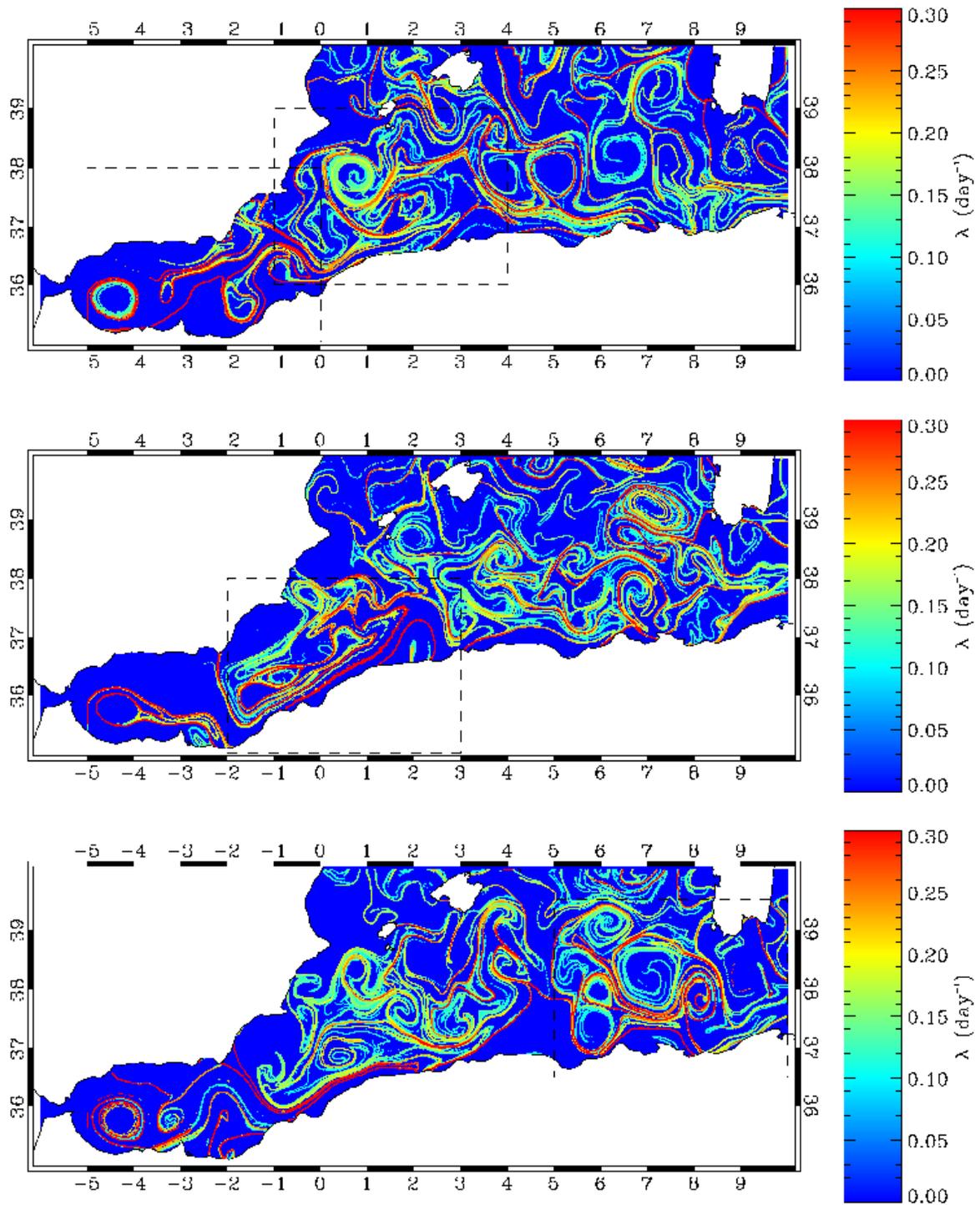} \\
\end{center}
\caption{FSLE ($\lambda$) computed from sea-surface height fields corresponding
(from top to bottom) to July 9, 2003; April 7, 2004;
June 30, 2004. The dashed boxes represent the regions over which
we perform a comparison with SST (see Fig.~\ref{fig:sst}). \label{fig:fsle}}
\end{figure}

Figure \ref{fig:okw} shows the spatial distribution of OW, which
appears as a set of vorticity-dominated ($W<0$, blue) regions
surrounded by strain-dominated lobular structures ($W>0$, red)
embedded in a background field of small values of $W$. On the
other hand, Fig.~\ref{fig:fsle} shows the spatial distribution
of FSLEs which has very different patterns characterized by a
tangle of lines. These are the locations were FSLE are large,
approximating unstable manifolds and corresponding to transport
barriers embedded in a background field with $\lambda\simeq 0$.
The unstable manifolds arise from fluid stretching and they are
therefore Lagrangian analogs of the deformation OW regions.

A first important difference in daily maps of OW and FSLE values is the presence
of eddies in the first and of filaments in the second. In particular, in the
FSLE map there are no enclosed regions and in some cases (e.g.  for the eddies
over the Balearic abyssal plain), the same manifold connects multiple lobes,
spanning a region of several eddies. When comparing these patterns with the
stream-function of the flow, SSH in this case, it can be observed that in
general eddies localized by the OW parameter and the centers of the spiralling
FSLE lobes are in good agreement with the extrema of the altimetric field, as
one would expect. Across the Algerian current, the OW field identifies some
possibly spurious eddies that do not appear in either the FSLE or SSH. This is
due to the fact that both SSH and FSLE are not invariant under a transformation
of coordinates to a frame of reference moving at a constant velocity with
respect to the original (Galilean transformation) and therefore, eddies are
hidden or partially hidden by the presence of the Algerian current. Since OW is
Galilean invariant it is able to detect these eddies but it fails in the
detection of the Algerian current which appears as a coherent structure
characterized by manifolds (barriers) parallel to the altimetric streamlines  in
the FSLE picture.

Another key difference between both fields is linked to the
time-evolution of coherent structures. From Figs. \ref{fig:okw}
and \ref{fig:fsle} it can be seen that eddies displaying a similar
size and intensity in the OW map may show very different features
in the FSLE map.  This is the case for instance for the western
Alboran eddy and some of the mesoscale eddies in the Algerian
basin (especially during July 30, 2004). For the Alboran vortex,
the lobe spiral is very tight, almost resembling the concentric
altimetry lines. For the eddies in the Algerian basin, the lobes
are loose, much less localized, and discordant with the altimetric
contours. It is interesting to notice that previous studies
\citep[e.g.][and references therein]{Isern.2006} have shown that
vortices in the Algerian basin propagate at velocities of the
order of 5 $km$ $day^{-1}$  in contrast to the western Alboran
eddy, which is almost stationary. In the ideal case of a
time-independent velocity field, particle trajectories, as well
as the set of lines in which Lyapunov exponents are organized (which
approximate material lines), coincide with altimetric isolines.
For the case of an eddy, they appear as concentric, closed
manifolds. In a time-dependent flow, the identity between
trajectories and altimetric isolines is lost and the differences
quantify the time variability. The material lines resemble
concentric circles for slowly evolving persistent structures like
the Alboran gyres and assume a complex shape in the case of more
dynamically active eddies, like the eddies along the Algerian
current and some of the eddies over the Algerian basin. This
phenomenon has an important effect on the transport properties.
Due to the fact that material lines act as transport barriers,
eddies with a low time variability and concentric unstable
manifolds have a smaller water mass exchange with the surrounding
compared to dynamical active vortices (see \citet{Lehahn.2007}
for a discussion of the eddy time variability in connection with
phytoplankton pattern formation).

A third important difference between the OW and FSLE approaches is
the spatial scale of the structures detected. The OW parameter is
bounded by the resolution of the altimetric data (1/8$^o$) and the
low-pass filters applied during the construction of altimetric
maps. This limitation does not hold for the FSLE that is based on
trajectory calculations, whose length scale is a combination of
both space and time variability of the velocity field. Lobes and
filaments below the altimetric resolution appear, especially in the
more dynamically active regions, like the fast evolving eddies
formed downstream of the Algerian current or the lobes south of
the Balearic islands. Very thin filaments are also associated with
the Algerian current.

\subsection{Detection of tracer fronts}

\subsubsection{Synthetic tracer}

\begin{figure} \begin{center}
\includegraphics[width=39pc]{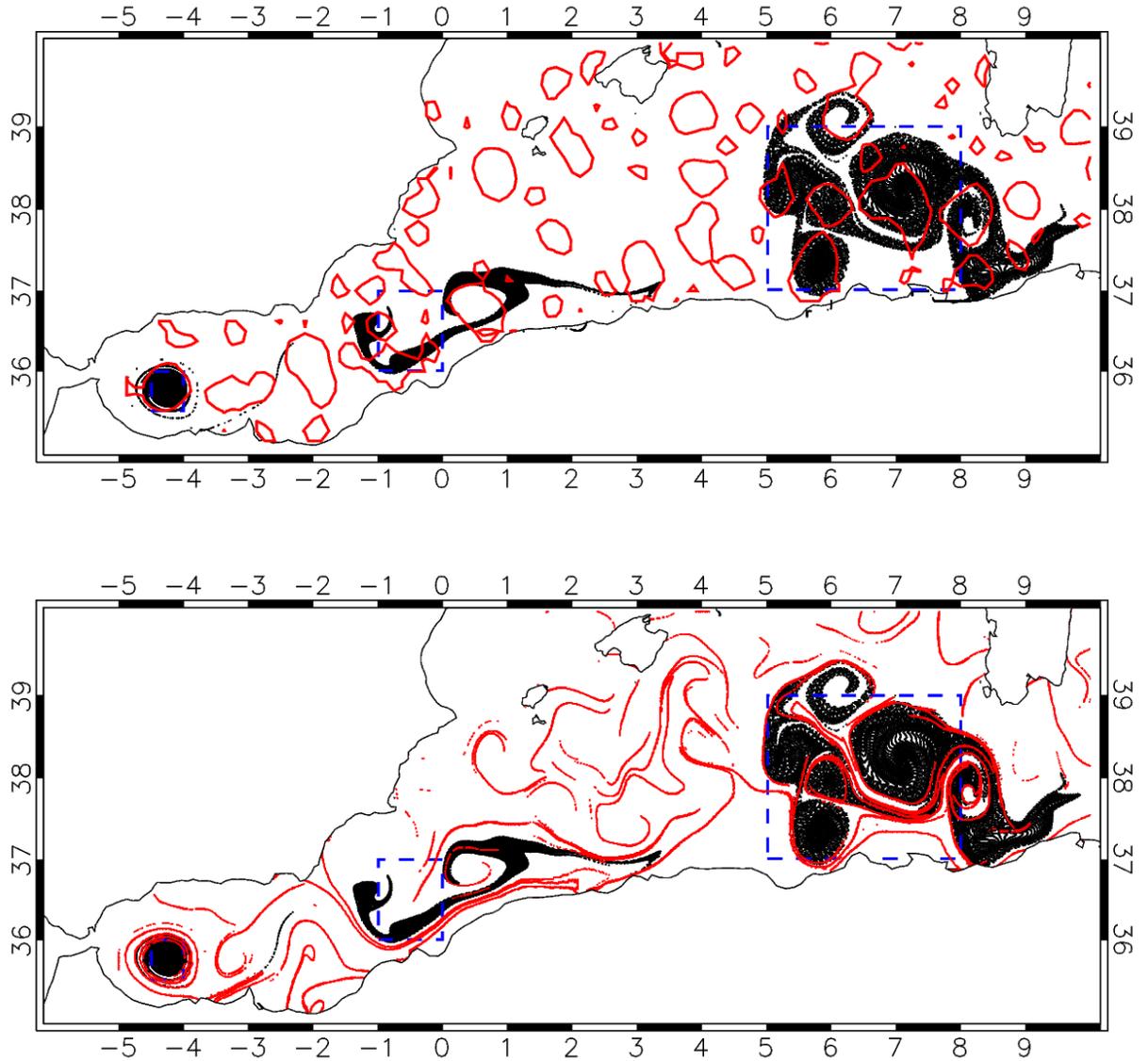} \end{center}
\caption{Distribution of tracer particles released over the Algerian current,
the western Alboran eddy, and the easternmost part of the Algerian basin (dotted
boxes). Tracers in the Algerian current are advected for one week, and the other
two sets are advected for two weeks. The release dates are chosen to have for
the three cases the coinciding final time (June 30, 2004) which is shown in the
figure. Zero-lines of OW corresponding to the final day are overimposed (top) as
well as the line-shaped regions where the backwards FSLE values are larger than 0.2
days$^{-1}$ (bottom), approximating unstable manifolds of the chaotic flow.
\label{fig:tracer}} \end{figure}

First, we test the ability of OW and FSLE to characterize tracer
distribution in an ideal but realistic situation: three sets of
particles are advected by the velocity field estimated from
altimetry. To this end we put three sets of particles distributed
on a square grid centered over three different dynamical
structures: the slowly evolving Alboran eddy, the Algerian
current, and the strongly interacting eddies in the easternmost
part of the Algerian basin (dashed boxes in
Fig.~\ref{fig:tracer}). Particles in the eddy regions have been
advected for two weeks. A shorter advection time of one week has
been used for the particles initially placed over the Algerian
current, due to the strongest velocity field in this region. The
advection time that we have chosen is such that the particles are
kept close to the dynamical structures we want to study: a larger
advection time does not change the results discussed, but increases
the dispersal of the particles over several structures.

The tracer released over the west Alboran eddy shows a regular, circular pattern
well correlated with both the OW and the FSLE maps. This pattern differs from
the more deformed distribution for the tracer released over the Algerian basin.
In this region, the tracer appears spread on several eddies connected through
thin filamentary structures. The boundaries detected by the OW parameter (we
plot in Fig. \ref{fig:tracer}, top, the lines $W=0$) provide an approximate
picture, often underestimating the size of the eddy cores and providing no
information of the patterns followed by the tracer exchanged from one eddy to
another. In contrast, the FSLE lines of intense stretching reproduce with
remarkable accuracy the tracer eddy boundaries, as well as the tangle of
spiralling filaments that connect them. This is seen in Fig. \ref{fig:tracer}
(bottom) where we plot the regions where the FSLE has values larger than
$0.2\;day^{-1}$. As discussed above, these regions are essentially
one-dimensional lines. They behave as material unstable manifolds of the
advecting flow: they are almost perfectly located along the tracer boundary.
Small deviations may be attributed to different reasons: (i) a tracer front
approaches the manifold exponentially fast, so that a residual distance remains
for a finite time of integration; (ii) only the manifolds with an intensity
larger than $0.2\;day^{-1}$ have been plotted and other weaker lines may also
act as transient transport barriers. The residual distance from a manifold can
be estimated by Eq.~\ref{eq:relax}: considering an initial tracer to manifold
distance of about 100 $km$, an exponent of $0.2-0.3\;day^{-1}$ for the manifold
(Fig.~\ref{fig:fsle}), and an advection time of 15 $days$, we get a separation
between the manifold and the tracer front of a few $km$. Note that the thin
filament at 7$^o$W, 37.5$^o$N is below the altimetric resolution, and appears in
both the tracer advection and the FSLE map.

Thin filaments also appear for the case of the tracer released over the Algerian
current. The OW is designed to detect vortices and therefore its use for barrier
detection along the Algerian current is, strictly speaking, improper.
Nevertheless, the gyre due to the mesoscale eddy located at 0.5$^o$W, 36.5$^o$N
is correctly predicted. Not surprisingly, features due to smaller and rapidly
evolving eddies (like the ones located along the current), as well as the
barrier effect due to the current itself, are not detected. Interestingly, an OW
signal appears in correspondence to the Almeria-Oran front (well visible in the
southern boundary of the tracer), probably as a signature of the secondary
Alboran eddy. Such a signature is composed of broken and irregular structures,
but nevertheless is in phase with the southern front of the tracer. The OW does
not provide any indication of the filament intruding the Algerian basin. In
contrast with the lobular structures detected for eddies, the FSLE shows for the
Algerian current meandering lines that follow the African coast. Such lines are
in almost perfect agreement with the tracer distribution, well indicating the
region of intrusion in the Algerian basin. A spiralling lobe at the location of
the eddy also detected by the OW method at 0.5$^o$W, 36.5$^o$N is correctly
localized. The Almeria-Oran front is well detected by a manifold that follows
uninterrupted the Algerian current, marking a transport barrier that confines
the water masses coming from the Alboran sea to an isolated tongue along the
African coast. In agreement with this picture, the tracer has no intrusion with
such coastal water, being initialized northern to such a manifold.

\subsubsection{An observed tracer: SST}

\begin{figure}
\begin{center}
\includegraphics[width=39pc]{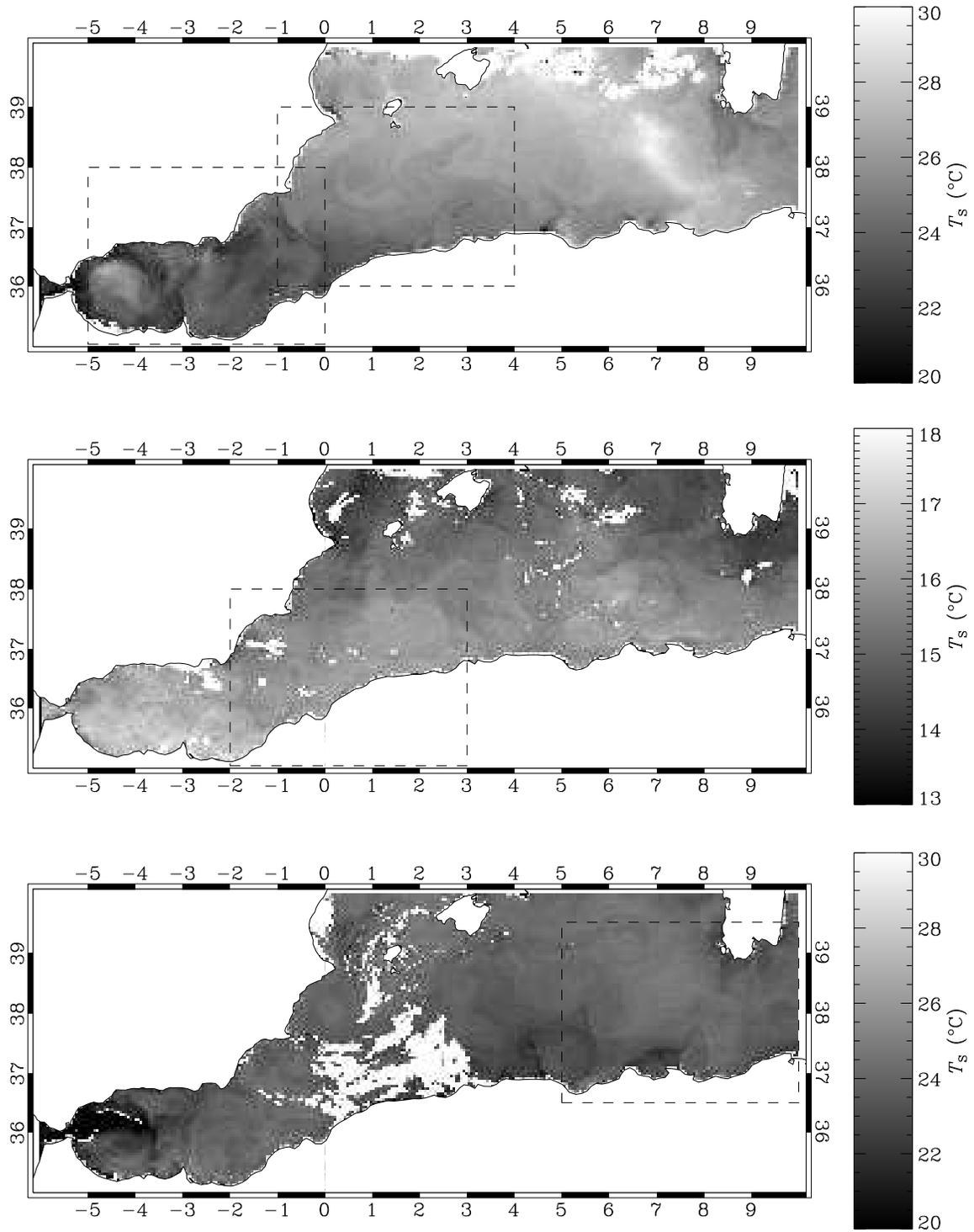} \\
\end{center}
\caption{General views of the sea-surface temperature fields
corresponding (from top to bottom) to July 9, 2003; April 7, 2004;
June 30, 2004.\label{fig:sst}}
\end{figure}

\begin{figure}
\begin{center}
\includegraphics[width=29pc]{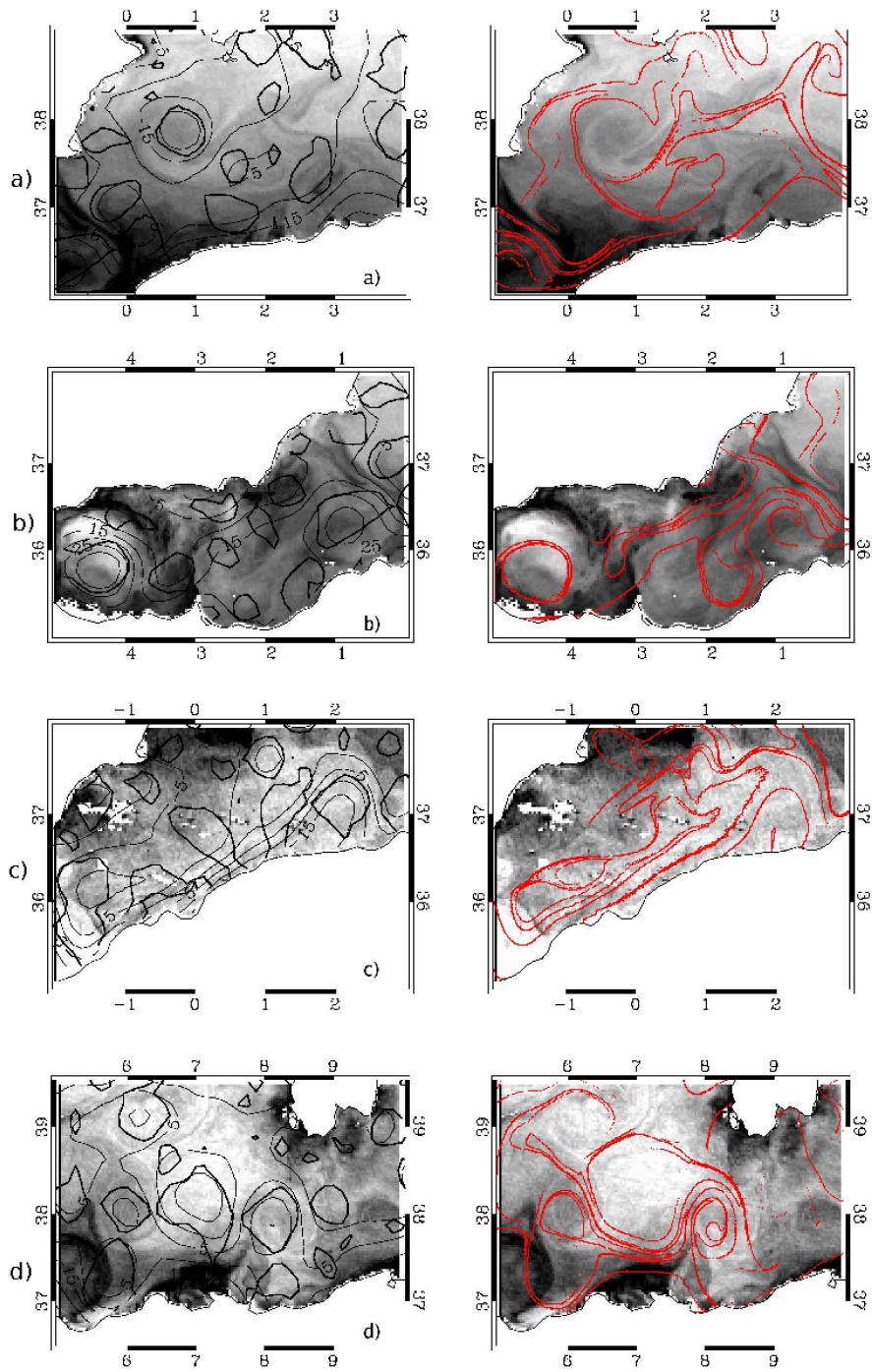} \\
\end{center}
\caption{Comparison of OW, FSLE and temperature distribution corresponding to the regions
shown in figures \ref{fig:okw}-\ref{fig:sst}. Left column: SST
(coded in grey levels) with SSH (thin black line) and lines of
zero OW (thick black line). Right column: SST (in grey levels)
with regions where FSLE is greater than 0.1 day$^{-1}$ (red
line-shaped regions). From top to bottom the dates are: July 9,
2003 (a and b); April 7, 2004 (c) and July 2, 2004 (d).
\label{fig:tracers}}
\end{figure}

Figure~\ref{fig:sst} depicts the
temperature distribution for the three days selected and
Fig.~\ref{fig:tracers} shows some zooms corresponding to the
dashed squares in Figs. \ref{fig:okw}, \ref{fig:fsle} and
\ref{fig:sst}.

Figure \ref{fig:tracers}a shows the situation of a relatively
isolated westward propagating Algerian vortex. The center of the
eddy is well located by the OW parameter, but the strongest SST
gradients are beyond the outside of the $W=0$ which identify the vortex
core. FSLE lines reproduce the two-lobe structure of the SST
positive anomaly contered  in 1E, 37.5N, although the size is
overestimated. Furthermore, south of this vortex there is a
filament of colder waters that approximately follows the transport
barriers depicted by FSLE. However, this example also shows one of
the limitations of the approach: FSLE and OW strongly depend on
the quality of altimetric maps. A coastal eddy in the Algerian
coast is clearly observable in the SST images but is not properly
captured by SSH, and therefore partially missed by OW and FSLE.

Figure~\ref{fig:tracers}b depicts the Alboran sea in spring (April
7, 2004). In its westernmost part, close to the Gibraltar strait,
a water mass of warm waters, surrounded by colder waters, is
trapped within the western Alboran gyre. Both the OW and the FSLE
maps show a regular and circular barrier although both seem to
underestimate the radius of the object, maybe due to the location
of the vortex in altimetric data. The Almeria-Oran front on the
eastern part of the image is observable in both SST and FSLE but
not in the OW field. Proceeding to west a large eddy attached to
the coast is quite properly identified in all fields: SST, FSLE,
SSH and OW. However, in the middle of the image, where the eastern
Alboran eddy is usually located there is a poor coincidence
between patterns calculated from altimetry and the SST image. In
particular, the SST fronts observed at 3$^\circ$W are almost
perpendicular to the FSLE lines there.

Figure \ref{fig:tracers}c shows an example of the strong signature
of the Algerian current. In contrast with the previous case, the
FSLE map reproduces with great accuracy the SST distribution. The
dynamical barriers due to the presence of the jet along the coast
are parallel to the SST isolines. In analogy to what was observed
for the synthetic tracer, the jet has no signature on the OW
parameter. On the other hand, several mesoscale eddy boundaries
with no effect on the SST pattern also appear in the OW image.
Centered at 37$^\circ$N 2$^\circ$E there is a large coastal eddy
probably generated by the destabilization of the Algerian current.
This eddy is clearly identified in all fields. However, the most
remarkable pattern is the deflection of coastal waters from coast
to the open sea due to the presence of this eddy, which is
observable in FSLE as well as in SST. Note also that water masses
in the inner part of the vortex are bounded by transport barriers
and therefore trapped within the eddy.

Finally, Fig. \ref{fig:tracers}d shows the easternmost part of
the Algerian basin where several eddies strongly interact. Notice
that this is one of the situations analyzed using the advection of
ideal particles in the previous section. As expected, the match
between real SST data and the FSLE is as good as before. The
manifold tangle observed in the FSLE map can also be seen in the
SST image and the cores of the eddies on which these manifolds
wind are identified also by the OW parameter. An isolated region
along the Algerian coast (7E, 37N), appearing in the SST image as
a strong cold anomaly, is also fairly well identified by FSLE and
not by the OW parameter.

\subsection{Time averaged fields}

In order to generalize the comparison performed on instantaneous cases, we focus
now on the relationships between spatial field distributions averaged over time.
To this end OW and FSLE have been computed for 10 years of data (1994-2004). The
OW parameter is calculated for each altimetric image (at one week time
resolution) while an FSLE map is generated each two days.

\subsubsection{Distribution of barrier-type lines}

\begin{figure}
\begin{center}
\includegraphics[width=39pc]{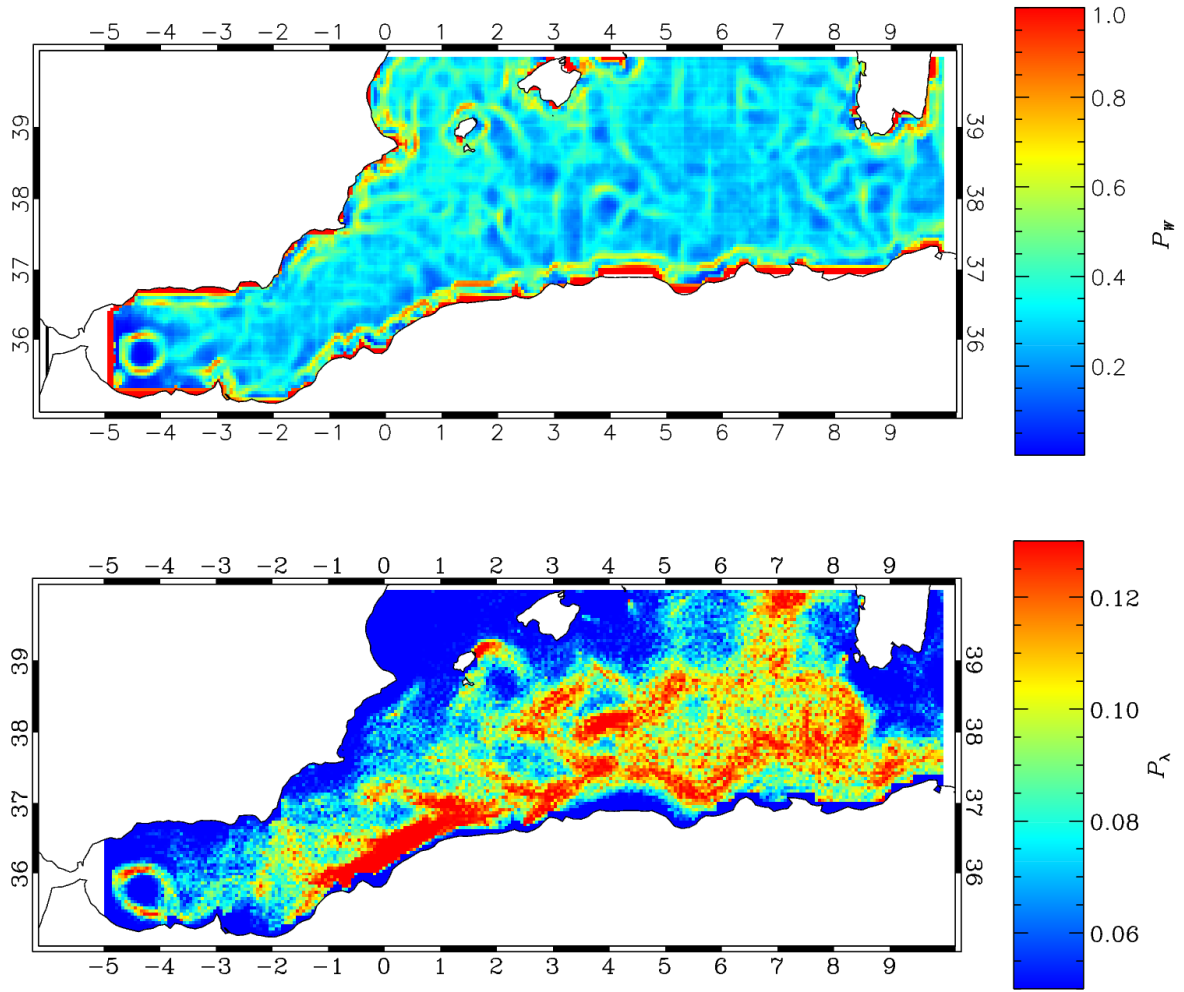} \\
\end{center}
\caption{Fraction of time during which each spatial
point is visited by the lines $W=0$ ($P_W$, upper panel) and by
the ridges (local maxima along some direction) of FSLE
($P_\lambda$, lower panel) for the period 1994-2004.
\label{fig:meandensityfsleow}}
\end{figure}

First, we estimate the fraction of time during which each spatial point is
visited by barrier-type lines. By barrier-type lines we mean lines (ridges made
of local maxima in FSLE and the $W=0$ isolines in OW) which could be
interpreted, at least during some short time, as a transport barrier. To this
end we count how many times each grid point pertains to one of such lines and
divide by the total number of observations for that point (the length of the
time-series). Figure \ref{fig:meandensityfsleow} shows these local probabilities
of having a transport barrier estimated using OW and FSLE (and denoted by $P_W$
and $P_\lambda$, respectively). In agreement to what we found in the previous
sections, we observe that the pattern corresponding to OW parameter has a more
regular, patchy structure due to the presence of Eulerian eddies, and no clear
signature for features that are not dependent on individual vortices, like the
Algerian current which appears instead in the FSLE map. This suggests the idea
that features that are common in both maps might be associated to long-lived
non-propagating vortices. The most evident example is the western Alboran eddy.
Other examples are the pattern observed around longitude 4$^\circ$E and
38$^\circ$N which is associated to the region of vortex detachment from coast
discussed in previous papers \citep{Isern.2006}, the pattern observed east of
Eivissa island (1.5$^\circ$E and 38.5$^\circ$N) or the Almeria-Oran front.

\begin{figure}
\begin{center}
\includegraphics[width=39pc]{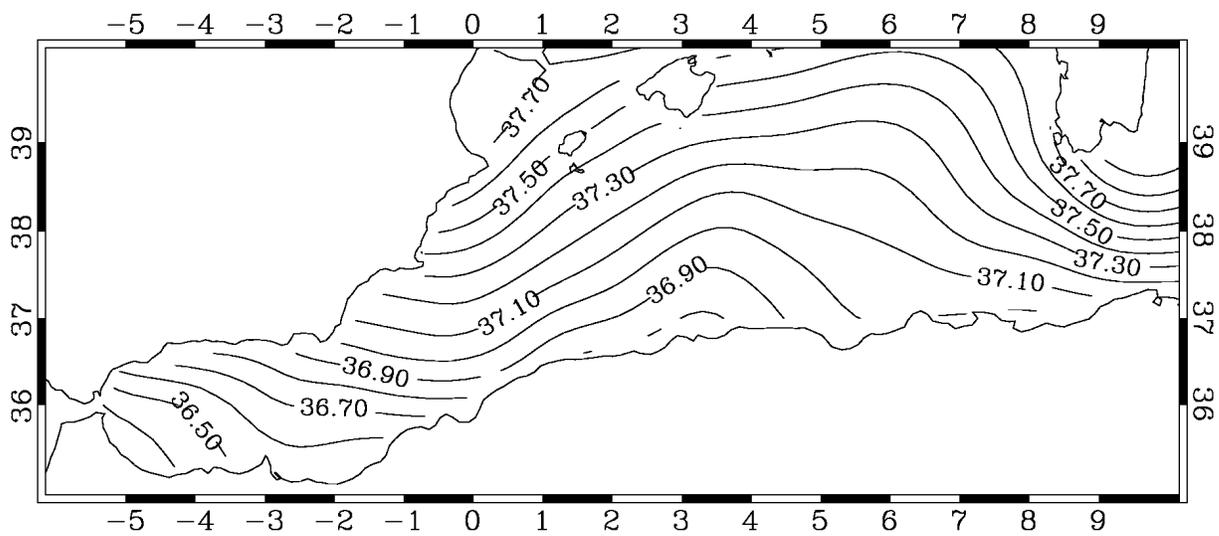} \\
\end{center}
\caption{Climatological distribution of salinity in the area of study at 5 m depth from
the MEDATLAS-II data set. \label{sal}}
\end{figure}

After observing that OW and FSLE provides very different
estimations of the propensity of the points to belong to a
barrier-like line, the next question is to identify which one is
closer to give a true barrier intensity and location. As outlined
in the introduction, a characteristic of the incoming Atlantic
waters with respect to the resident Mediterranean ones is their
lower salinity. If there are significant barriers to the spread of
this surface water into the Mediterranean, climatological
distributions of salinity should locate them. Figure \ref{sal}
shows the climatological salinity obtained from MEDATLAS-II data
set. When comparing Figs. \ref{fig:meandensityfsleow} and
\ref{sal} it is evident that the estimation computed from FSLE
provides a better picture, although the smoothness of the
climatological salinity makes difficult the comparison. On
Alboran, salinity increases eastwards. Close to the Almeria-Oran
front the isohalines are almost aligned width the front, as
depicted by FSLE. Eastwards, the intrusion of fresh waters along
the coast following the Algerian current and the associated
intense northwards gradient matches quite well the patterns
observed in Fig. \ref{fig:meandensityfsleow} (bottom).
Proceeding to the east, at the entrance of the Sardinia channel
FSLE barrier density is concentrated close to the coast, in
correspondence with low salinity waters being also confined to
this area. In the middle of the basin, the probability of having
transport barriers is more homogeneous due to the propagation of
Algerian eddies and consequently climatological salinity is more
uniform.

\subsubsection{Time averages of FSLE and other Eulerian diagnostics}

Results in previous sections suggest that high FSLEs are preferentially located
on energetic structures like at the eddy periphery or along jets. This suggest a
possible correlation between the FSLEs and other Eulerian diagnostics different
from OW, such as the eddy kinetic energy (EKE) or the strain. In fact, in a
recent study of the Tasman sea, it has been shown that averages of both the EKE
and the strain have a compact relationship with averages of finite-time Lyapunov
exponents \citep{Waugh.2006}. Although such relationships are valid only on time
average, and therefore cannot be used for the detection of instantaneous, local
transport barriers, the correlation between EKE, strain, and Lyapunov exponents
would provide a simple way to estimate Lyapunov exponent climatologies, as also
suggested in \citet{Waugh.2006}. Finite-time and finite-size Lyapunov exponents
only differ on the determination of the integration time (set a priori for
finite-time, given indirectly by prescribing the final separation of the
trajectories for finite-size). For this
reason, we tested the hypothesis above by computing for our data period time
averages of strain rate and of EKE.

\begin{figure}
\begin{center}
\includegraphics[width=39pc]{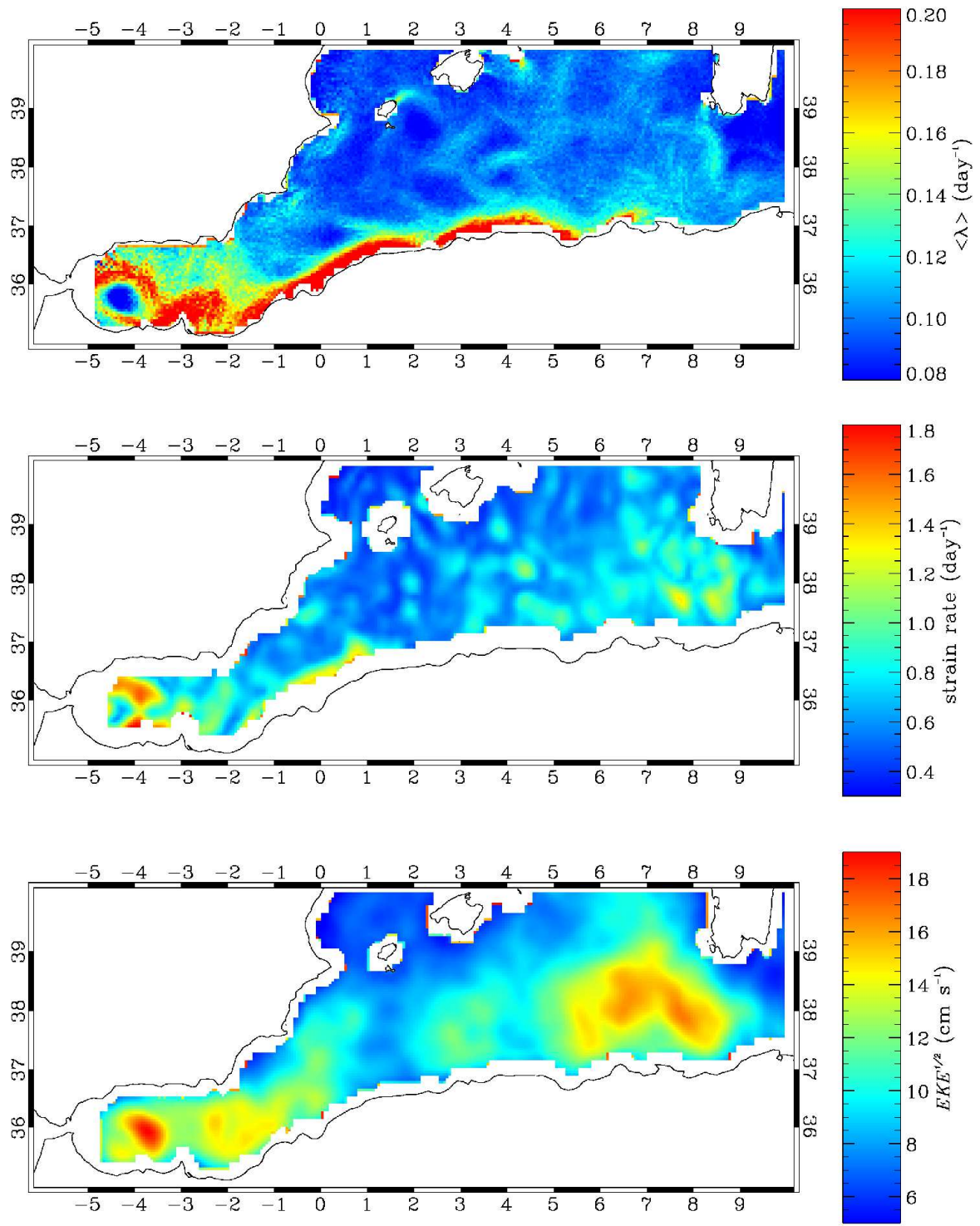} \\
\end{center}
\caption{Time-averaged FSLE, strain rate and eddy kinetic energy (1994-2004).
\label{fig:meanfsle}}
\end{figure}

Climatologies of FSLE, EKE, and strain rate are shown in Fig.~
\ref{fig:meanfsle}. Similar structures appear in regions containing persistent
objects, such as the Alboran gyre and the Algerian current. A large-scale
gradient is also observed, with more signal in the south than in the north of
the Algerian basin, and positive anomalies in the Sardinia channel. However,
there are more localized mesoscale anomalies in the FSLE field.

\begin{figure}
\begin{center}
\includegraphics[width=39pc]{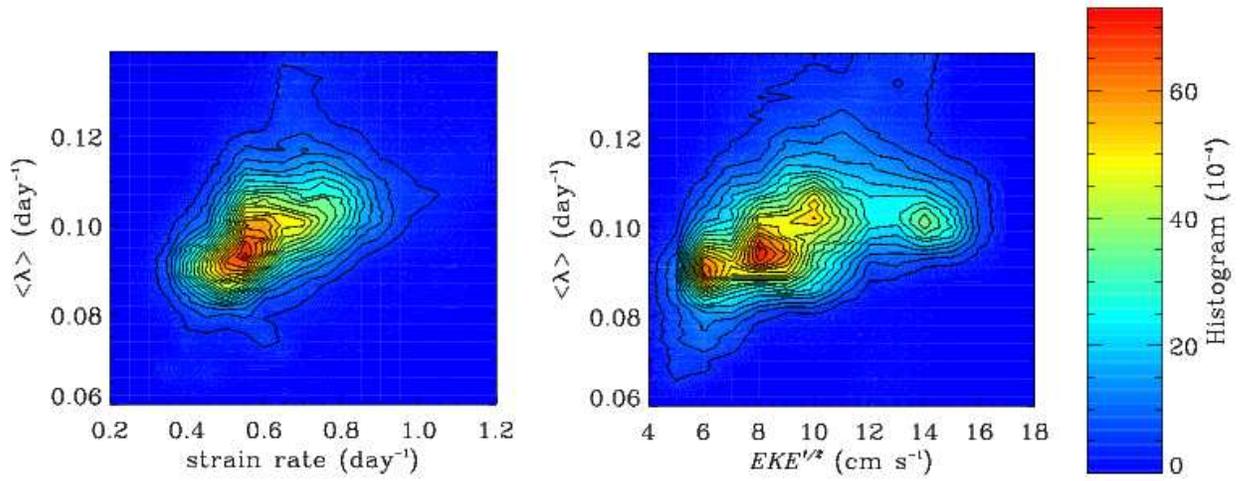}
\end{center}
\caption{Joint distributions of Finite-size Lyapunov exponents
vs. strain rate values (left),
and Finite-size Lyapunov exponents vs. eddy kinetic (right)
from the spatial distributions time-averaged over the period
1994-2004.
\label{fig:fslevssteke} }
\end{figure}

In Fig.~\ref{fig:fslevssteke}, joint distributions of FSLEs vs strain rate and
EKE are presented. Although theres is a general positive correlation among these
quantities, they show a much looser relationship than the one observed by
\citep{Waugh.2006} at lower resolution and on shorter integration times. The lack of compact relationships on the
scatter plots, as well as different distribution of anomalies among the various
diagnostics is a clear confirmation that the Lyapunov exponent provide a
complementary information in respect to Eulerian diagnostics even in a
climatological sense, when the Lyapunov calculation is done at high resolution.
The comparison with other Eulerian diagnostics (like the vorticity and the
vorticity gradient, that is sometimes also used for transport barrier detection,
see for instance \citep{Paparella.1997}) confirm these observations.

\section{Summary and discussion}

In this work we have exploited the rich mesoscale activity of the Algerian basin
for comparing Eulerian and Lagrangian diagnostics of transport under various
conditions. This comparison unveiled two coexisting pictures: an Eulerian view,
dominated by mesoscale eddies, and a Lagrangian view, characterized by
interconnected lobular structures and submesoscale filaments. The first picture
describes the eddies that populate the mesoscale turbulence; the second one
describes the spatial structures of tracers advected by these eddies. Eddy cores
are usually identified by setting a threshold ($W=0$) on the OW parameter, and
this $W=0$ line may be a rough candidate to a `transport barrier' since eddy
contents are seen to remain coherent in several situations. More precise
candidates to transport barriers  are obtained from local maxima (ridges) of the
FSLE computed backward in time.

As diagnostics of transport barriers, both techniques are based on heuristic
arguments related to the time variability of the velocity field. The OW use is
based on the consideration that for time-invariant fields, Lagrangian and
Eulerian barriers coincide. For this reason, with the OW method no structure
typical of time varying fields like filaments and lobular patterns can be
detected but nevertheless remarkably good results can be obtained for stationary
eddies, in our case the first Alboran gyre and some of the most persistent
eddies over the Balearic abyssal plain. The FSLE calculation goes one step
further, assuming quasi stationarity not for the velocity field, but for the
lines of maximal divergence (unstable manifolds of hyperbolic points). When this
time-scale separation between the Lagrangian structures and the advected tracer
holds, the tracer front relax exponentially fast over the ridges of
large Lyapunov exponents that therefore mark the front position. Another advantage of the
FSLE method is that the effect of both eddies and jets are remarkably dealt
with.

We have tested the barrier candidates against a synthetic tracer advected by the
geostrophic field. This test sets an upper limit on the confidence on FSLE and
OW assumptions, since it excludes the effect of any factor acting on the tracer
except the geostrophic velocity field. At the meso- or larger scales, the OW was
found to locate correctly (with errors of tens of km) tracer patches when they
were well confined inside eddies. Fluid parcels intruding eddies's along thin
filaments are not captured however. Such filaments, as well as the submesoscale
structure of tracer fronts, were found from the FSLE calculations.

In fact, the advection of a synthetic tracer can be considered itself as a
direct, Lagrangian diagnostic of transport barriers. This approach however has
some limitations. Lagrangian structures are sampled unevenly, since the tracer
tends to spend longer times in regions with low velocities. For this reason,
even strong barriers may require a carefully choice of the initial conditions
and of the integration time. An example of this appears in
Fig.~\ref{fig:tracer}, where part of the Almerian-Oran front is not entirely
visible and would have required to probe the velocity field with a larger number
of tracer blobs. When doing the tracer experiment of Fig.~\ref{fig:tracer} we
also observed that, not surprisingly, the tracer has a long residence time over
the Balearic Abyssal Plain and therefore for long integration times
most of the time the tracer shades the barrier of this region only,
independently of its initial condition. On the other hand, a reduction of the 
integration time increases the gap between the tracer front and the actual
barrier position, since the tracer front has a shorter time to relax to it (see
the discussion of Eq.~\ref{eq:relax}). In fact, an
attempt to overcome these limitations requires the use of backward trajectories
(for which barriers are regions of maximal separation) and of variable 
integration times, basically yielding a FSLE-type calculation. For this reason,
the FSLE can be considered an optimised way of advecting a passive tracer for
the detection of transport structures. Note also that, besides guaranteeing a
uniform sampling of the velocity field and the precise localization of transport
barriers, the FSLE also provides the information of the intensity of the barrier
at the same cost of tracer advection.

The representation of advection in the Algerian basin however is based on
altimetry and therefore is limited by the lack of representation of any
non-geostrophic component of the velocity field as well as unresolved or poorly
resolved structures due to space and time resolution \citep{Pascual.2006}, and
noise. The comparison with SST images allows to verify that such effects are not
strong enough to spoil the FSLE usefulness in locating Lagrangian features. Note
that the filament tangle connecting different eddies as well as the spiral
structures inside eddy cores cannot appear in any way in a time-independent
field (i.e., by using a snapshot of the velocity field), where one can show that
orbits correspond to altimetric contours, and eddy barriers to closed contours.
Nevertheless, the slow time variability of a few altimetric frames is enough to
generate Lagrangian structures which correspond well to SST images. We also
expected an effect of altimetric errors especially for the structures close to
the coast, but with the exception of the position of the first gyre in the
Alboran sea and in spite of the chaotic nature of the geostrophic advection, the
localization of features from Lyapunov exponent maps appears quite reliable.

\begin{figure}
\begin{center}
\includegraphics[width=39pc]{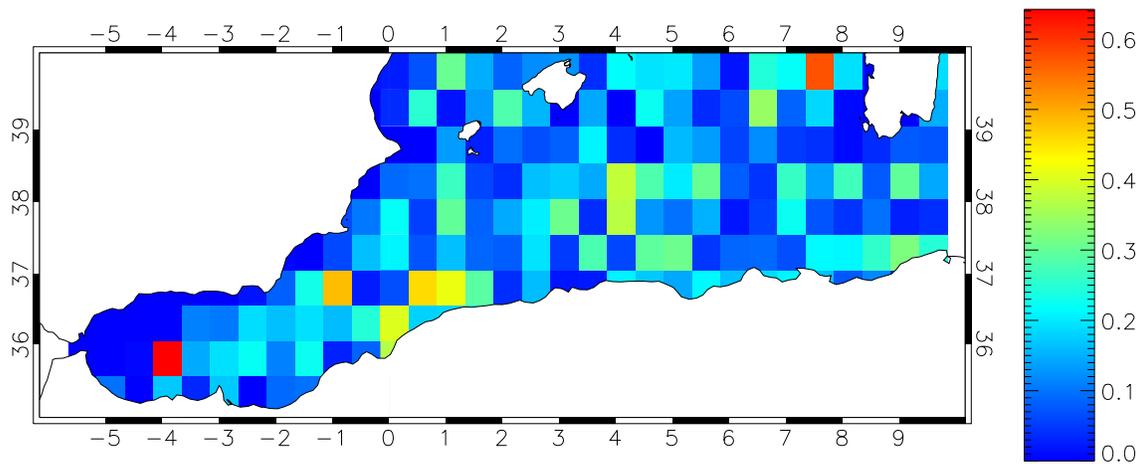} \\
\end{center}
\caption{Calculation of the FSLE for July 9, 2003 with initial separation
$\delta_0=0.5^o$. Compare with Fig.~\ref{fig:fsle} a. At this resolution, Lyapunov exponents
cannot be linked to tracer filaments.
\label{fig:fsle05}}
\end{figure}

Climatologies of several Eulerian diagnostics and either barrier densities or
average values of Lyapunov exponents also showed distinct different patterns.
Given the fact that the main differences between Lyapunov and Eulerian
diagnostics are related to small scale filaments and in the temporal variability
of mesoscale turbulence, it is not difficult to expect different relation
between Eulerian diagnostics and Lyapunov exponents computed at much larger
scale and on shorter time periods. This is the case of the 0.5 deg. and 15 days
calculation of in \citet{Waugh.2006} (resp. 50 times larger and 3 times shorter
than in our case). We illustrate this in Fig.~\ref{fig:fsle05}, where the
Lyapunov exponents have been computed with initial separations of 0.5 deg. for
the same day of Fig.~\ref{fig:fsle}a. Indeed, at the resolution of the
Mediterranean altimetry product (1/8 deg.) not even the patterns of two Eulerian
diagnostics like the strain and the eddy kinetic energy appear to be correlated
(Fig.~\ref{fig:meanfsle} b and c). An interesting exercise would be the
calculation of correlation between eddy kinetic energy and Lyapunov exponents
computed at various scales, in order to see whether there exists a critical
lengthscale or a smooth transition. In fact, the finite-size Lyapunov exponent
has been introduced in turbulence theory with the aim of probing a velocity
field at various scales. General arguments \citep{Aurell.1997} suggest the
presence of a cross-over scale in the Lyapunov calculation separating an
exponential regime from other diffusive regimes at the scale of the smallest resolved
eddy. For the case of altimetry, this cross-over can be expected to appear at
spatial scales of the order of 1 deg.

Note that the Lyapunov exponent is not an advanced replacement of Eulerian
diagnostics, but a complementary tool incorporating tracer information. For instance, the
association of lobes with eddies that in some cases we observed only holds for
stationary eddies. If used systematically as a Lagrangian eddy detection, this
approach would lead to detect spurious, small-scale spatial signal, that in fact
comes from the temporal variability of large-scale eddies. The OW and FSLE
method are also complementary from a technical viewpoint. The OW method is a
differential technique. It is thus affected by small-scale noise and especially
by discontinuities. These kind of features indeed can appear in the presence of
eddies with a size close to the spatial resolution. On the contrary, the FSLEs
are computed by integrating particle trajectories and thus are not strongly
affected by local noise or discontinuities.  The complementarity of the
information detected by FSLE in respect to the OW parameter and other Eulerian
diagnostics is also emphasised in the climatologies of
Fig.~\ref{fig:meandensityfsleow} that showed how the differences detected in
single day analyses are indeed a persistent feature.

The main drawbacks of the FSLEs are a relatively increased complexity in the
calculation (with respect to the OW) and especially the need of extended
velocity data sets, both in time and in space. This can be a problem if the
observational region is limited (either in time or space), since some
trajectories may go out from the boundaries  (this is in fact at the basis of
other methods used to characterize flow structures \citep{Schneider.2005}).

The OW on the contrary can be calculated from a snapshot of the
velocity field. Indeed, its easiness allows to compute it even
from in situ data
\citep[e.g.][]{Stocker.2003,Testor.2005,Pallas.2005} and then
obtain a first approach to the flow topology and its effect on the
transport. Furthermore, although OW is a differential technique
and therefore more sensible to velocity errors or discontinuities,
it is possible to approach it using kinetic energy at least for
the identification of vortex core edges
\citep{Paparella.1997,Isern.2004}. If the vortex is
assumed to be close to axial symmetry then the OW formula can be
written as
\begin{equation}
W\propto\frac{1}{r}\left(\frac{\partial E(r)}{\partial r}\right),
\end{equation}
where $r$ is the radius from the vortex center and $E(r)$ its
density of kinetic energy at that distance from the center. This
implies that the line of $W=0$ will correspond to a local maximum
of energy.

The comparison with SST and synthetic tracer patterns also shows that the errors
in the OW method and the FSLE technique are of very different nature. The OW
fails with both the real and synthetic tracer mainly for detection of the
filamentation process, due to its intrinsic lack of representation of the time
variability of the velocity field. In contrast, the FSLE is limited not by its
assumption of time-scale separation between barrier and tracer, but because of
the limits of the altimetric data. This is clearly shown by the fact that the
FSLE performs almost ideally for the synthetic case while only providing a more
qualitative estimations of SST fronts. In perspective, envisaging the
availability of assimilation and observational velocities field of increased
quality and spatiotemporal resolution, one can expect a decrease in effectiveness
of the OW technique in favour of the FSLE (or any other Lagrangian technique) at
least for the detection of (sub-)mesoscale transport barriers.

\section{Conclusions}

The advantages of a Lagrangian tool when compared to Eulerian diagnostics in
capturing tracer structures would be trivial for model data. What is not trivial
is that altimetry data are amenable to such a comparison, given that the
differences arise in presence of tracer filaments, i.e., at a scale similar or
even smaller than the nominal resolution of altimetry data (and much smaller
than the cutoff in the altimetry energy spectrum). At this scale ageostrophic
components (unresolved by altimetry data) as well as SST active dynamics is
supposed to play an important role. The fact that neverthelss altimetry-derived
patterns approach SST structures means that these unresolved components of the
flow are either of weak intensity or act in phase with the horizontal stirring
\citep{Lehahn.2007}. The recent availability of realistic, submesoscale
resolving biogeochemical models (e.g. \citep{resplandy.2008}) should give the
possibility of exploring the role of unresolved ageostrophic components and of
tracer activity on filament detection.

The fact that altimetry data can be pushed at subgrid scales when re-analised by
Lagrangian tools, providing systematic information of filament location, opens
several interesting perspectives. For the case of high-resolution biogeochemical
measurements, the Lyapunov exponent provides an unpaired tool helping to
interpret the measurements in terms of the precise position of mesoscale and
sub-mesoscale fronts. The use of Lyapunov calculation can be even concieved in
real-time, for targetting frontal structures during in-situ surveys. This
possibility comes from the fact that the Lyapunov calculation only requires the
past history of the velocity field, and from the availability of altimetry
products with 1-2 weeks of delays (e.g. from AVISO).

A second interesting possibility is the use of the Lyapunov exponent together
with high-resolution tracer measurments for the validation of a velocity field.
An improved altimetry product (as what is expected from the recent launch of
Jason-2 and incoming altimetry satellites) should bring a velocity field at
better spatial and temporal resolution. The possibility of computing with
Lyapunov calculation the position of fronts and of comparing them with
high-resolution SST or chlorophyll images is one of the very few possibilities
for testing the ability of the new products of better representing mesoscale and
sub-mesoscale processes. This validation would be especially useful for choosing
the optimal combination of spatial and temporal resolution of surface current
products. A
similar approach can be used also for validating assimilation products and in
general for tuning the spatiotemporal resolution of circulation models.

Given the fact that transport barriers in the ocean
evolves on a much slower time scale than advected tracers, one can also
approximate the short-time advection of a released tracer as a relaxation
dynamics towards the stronger transport barriers and in this way estimate the
tracer ditribution in the near future. The most evident application of this
system would be, for example, the prediction of the evolution of pollutants.
Other Lagrangian techniques have already approached this problem
\citep{Lekien.2005}.


\section{Acknowledgments} 

This work is a contribution to the FISICOS (FIS2007-60327) project of the
Spanish MEC and PIF project OCEANTECH of the Spanish CSIC. FdO has
been supported by Marie-Curie grant 024717-DEMETRA and CNES. JIF has been
supported by Marie-Curie grant 041476-OCEAN3D. The altimeter products were produced by Ssalto/Duacs and distributed by
Aviso, with support from CNES (http://www.aviso.oceanobs.com). AVHRR  data were obtained from
the SAIDIN database at ICM  (http://ers.cmima.csic.es/saidin/). We acknowledge
the help received  from Marie-H\'el\'ene Rio, Alejandro Morales and Isabelle
Taupier-L\'etage.


\bibliographystyle{elsart-harv}
\bibliography{biblio_imagen}


\end{document}